\documentclass[aps,prb,reprint,showpacs]{revtex4-1}
\usepackage{graphicx}
\usepackage{epstopdf}
\newcommand{\et}{\textit{et al.} }
\usepackage[retainorgcmds]{IEEEtrantools}

\begin{document}
\newcommand{\gtwo}{$g=0.0496\pm0.0003$}
\newcommand{\xthree}{$x_3^{-1}=-0.201\pm0.015$}

\title{Nonlinear field-dependence and $f$-wave interactions in superfluid $^3$He}
\author{C.A. Collett}
\author{J. Pollanen}
\author{J.I.A. Li}
\author{W.J. Gannon}
\author{W.P. Halperin}
\affiliation{Department of Physics and Astronomy, Northwestern University, Evanston, IL 60208, USA.}
\date{\today}

\begin{abstract}
We present results of transverse acoustics studies in superfluid $^3$He-$B$ at fields up to 0.11 T.  Using acoustic cavity interferometry, we observe the Acoustic Faraday Effect for a transverse sound wave propagating along the magnetic field, and we measure Faraday rotations of the polarization as large as 1710$^\circ$.  We use these results to determine the Zeeman splitting of the Imaginary Squashing mode, an order parameter collective mode with total angular momentum $J=2$. We show that the pairing interaction in the $f$-wave channel is attractive at a pressure of $P=6$ bar. We also report nonlinear field dependence of the Faraday rotation at frequencies substantially above the mode frequency not accounted for in the theory of the transverse acoustic dispersion relation formulated for frequencies near the mode. Consequently, we have identified the region of validity of the theory allowing us to make corrections to the analysis of Faraday rotation experiments performed in earlier work.
\end{abstract}

\pacs{43.35.Lq, 67.30.H-, 74.20.Rp, 74.25.Ld}

\maketitle

\section{Introduction}
Superfluid $^3$He-$B$ is the only liquid known to support transverse sound. While first predicted in normal $^3$He by Landau,\cite{lan.57b} collisionless transverse sound was not realized until Moores and Sauls \cite{moo.93} showed that transverse mass currents couple to the Imaginary Squashing mode (ISQ), leading to propagation in $^3$He-$B$. After its discovery by Lee \et in 1999,\cite{lee.99} transverse sound has been exploited as a probe of the excitation spectrum of $^3$He-$B$, including a number of studies\cite{lee.99,dav.08a} which have used the Acoustic Faraday Effect to measure the Zeeman splitting of the ISQ from which the magnitude of $f$-wave pairing interactions in superfluid $^3$He was calculated. In the present work \cite{col.12a} we have extended those studies to much larger magnetic fields where we have observed Faraday rotation angles as large as 1710$^\circ$, entering regimes where nonlinear field effects play a role and sound frequencies are significantly higher than the ISQ frequency. Under these conditions we have found discrepancies with the theory,\cite{sau.82,sau.00a} which was formulated for sound frequencies near the mode. We have devised a phenomenological model that relates our results to the region of applicability of the theory. With this relation we have determined more precise values for both the Zeeman splitting and the $f$-wave pairing interactions than was previously possible, and we report our observation of nonlinear field effects on the ISQ.
\begin{figure}[b!]
\includegraphics[width=3.4in]{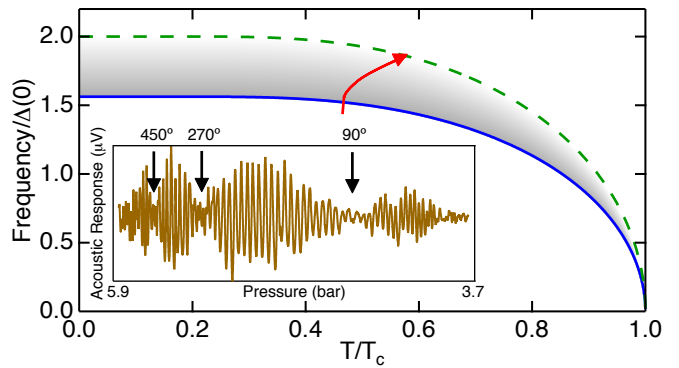}
\caption{\label{fig:EnDiag}Frequencies of the ISQ (blue line) and pair-breaking edge (green dashed line) as a function of reduced temperature. The grey shaded area is the region supporting transverse sound. The red arrow indicates the path of the sound frequency, normalized to the pressure dependent gap, during a typical decreasing pressure sweep. Inset: A representative acoustic response versus pressure measured between the ISQ and pair-breaking, taken on a path similar to that indicated by the red arrow, at $H=0.04$ T, with the visible minima, indicated by black arrows, corresponding to AFE rotations of $90^\circ$, $270^\circ$, and $450^\circ$.}
\end{figure}

Transverse sound provides a highly sensitive spectroscopy for the ISQ and its dependence on magnetic field. The frequency of this collective mode in zero field has been shown to be $\Omega_0\sim\sqrt{12/5}\Delta$,\cite{dav.06,dav.08a} where $\Delta$ is the weak-coupling-plus gap. \cite{rai.76} In a magnetic field $\Omega$ splits into five Zeeman sub-states of which only two, $m_J=\pm1$, couple to transverse sound.~\cite{moo.93} The theory of Moores and Sauls \cite{moo.93} shows that right circularly-polarized (RCP) and left circularly-polarized (LCP) sound couple to opposing $m_J$ states, causing acoustic circular birefringence of a propagating linearly polarized transverse sound wave. As the field strength increases, the difference between the velocities of RCP and LCP sound increase proportionately, resulting in a rotation of a linearly polarized acoustic wave. This Acoustic Faraday Effect (AFE) was first reported by Lee \textit{et al.},\cite{lee.99} providing proof that transverse sound is a robust propagating acoustic mode in superfluid $^3$He.

The coupling between transverse sound and the ISQ is described by the following dispersion relation, which holds in the limit that the acoustic frequency $\omega$ approaches the ISQ frequency $\Omega$:
\begin{equation}
\frac{\omega^2}{q^2v_F^2}=\Lambda_0+\Lambda_{2^-}\frac{\omega^2}{\omega^2-\Omega^2(T,P)-\frac{2}{5}q^2v_F^2}, \label{fulldisp}
\end{equation}
where $q$ is the wavevector and $v_F$ the Fermi velocity. The quasiparticle restoring force is $\Lambda_0=\frac{F_1^S}{15}(1-\lambda)(1+\frac{F_2^s}{5})/(1+\lambda\frac{F_2^s}{5})$, and $\Lambda_{2^-}=\frac{2F_1^s}{75}\lambda(1+\frac{F_2^s}{5})^2/(1+\lambda\frac{F_2^s}{5})$ is the superfluid coupling strength, where $F_1^s$ and $F_2^s$ are Landau parameters, and $\lambda$ is the Tsuneto function.\cite{sau.00a} Up to linear order in magnetic field, $H$, the splitting of the ISQ is expected\cite{sau.00a} to modify the denominator on the right-hand side of Eq. \ref{fulldisp} to be
\begin{equation}
\omega^2-\Omega_0^2-\frac{2}{5}q^2v_F^2-2m_Jg\gamma_{\text{eff}}H\omega. \label{linisq}
\end{equation}
Here $\gamma_{\text{eff}}$ is the effective gyromagnetic ratio of $^3$He and $g$ is the Land\'e $g$-factor of the ISQ. It should be noted that Eq. \ref{linisq} does not correspond to replacing $\Omega^2(T,P)$ in Eq. \ref{fulldisp} by its field dependent form $\Omega^2(T,P,H)$.

In an early study of the field dependence of the ISQ, Movshovich \et \cite{mov.88} made measurements up to $0.46$ T with longitudinal sound, which is strongly coupled to the ISQ. This coupling causes large extinction regions around the mode frequency, making it impossible to differentiate different Zeeman substates at low fields. In contrast to longitudinal sound, the transverse mode is weakly coupled to the ISQ and consequently has much higher spectral resolution and is more suitable for comparison with existing theory.

\begin{figure*}[t!]
\includegraphics[width=7in]{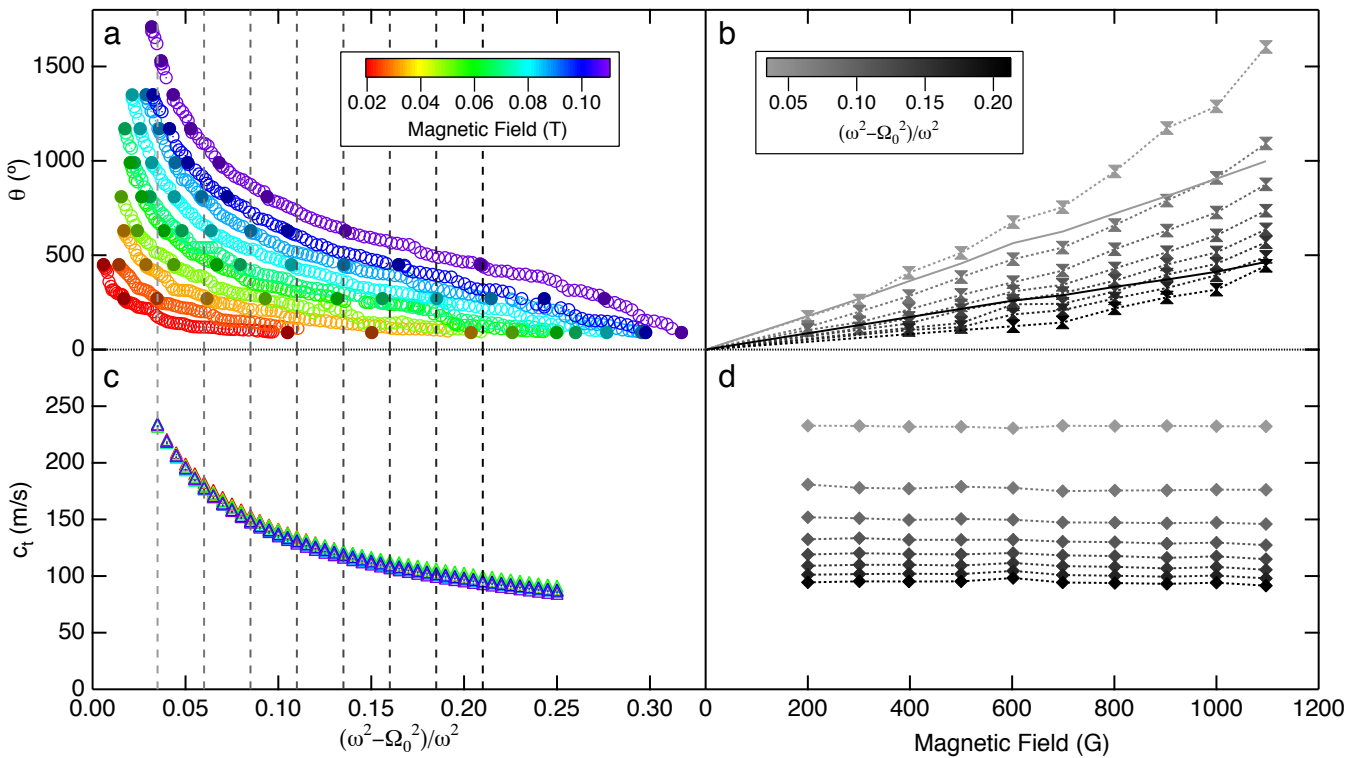}
\caption{\label{fig:data}AFE rotation angle, $\theta$, and sound velocity, $c_t$, data at all experimental fields, as a function of both shift and magnetic field. The vertical dashed lines in (a) and (c) indicate the shift values for which data is presented in (b) and (d). (a) Data for $\theta$ as a function of shift. The solid circles correspond to minima in the acoustic trace. (b) A subset of $\theta$ data at representative values of $(\omega^2-\Omega_0^2)/\omega^2$ as a function of field, taken at shifts from $(\omega^2-\Omega_0^2)/\omega^2=0.035$ to 0.21 at intervals of 0.025, demonstrating nonlinear behavior and a decrease in $\theta$ as a function of distance from the ISQ. The dotted lines directly join the data points, and are presented solely as guides to the eye. The solid lines indicate the $\theta$ values expected from the dispersion for linear splitting at constant $g$ at the lowest (light grey, 0.035) and highest (black, 0.21) shifts presented, exhibiting less than two thirds the decrease with field shown in the data. (c) Data for $c_t$ as a function of shift; the data points for different fields overlap at each shift value. (d) A subset of $c_t$ data at the same shift values as in (b), demonstrating a very slight field dependence of $c_t$.}
\end{figure*}

Previous transverse sound experiments\cite{lee.99,dav.08a} measured the $g$-factor at fields below $H\approx0.04$ T over a wide pressure range, observing purely linear field dependence. At the high fields used by Movshovich {\it et al.}~\cite{mov.88} quadratic effects were evident. Thus the intermediate field region $0.04\lesssim H\lesssim0.1$ T, where nonlinear field effects become significant, has remained relatively unexplored. In the present work we investigate both this field region and regions of frequency well above the ISQ frequency. This allows us to better determine the regime where the theory is valid, and correspondingly identify the low-field, linear Zeeman splitting and the corresponding $f$-wave pairing interactions in superfluid $^3$He.

In addition to the AFE, an applied magnetic field induces acoustic circular dichroism. The absorption coefficients of RCP and LCP sound depend on $m_J$, and in a field they have different values, causing one polarization to be attenuated more than the other. This has the effect of both flattening and shifting the Faraday rotation envelope. These effects are not significant in fields in the 0.1 T range, and so do not play a role for the field strengths used in our experiments.\cite{sau.12a}

\section{Experiments}
Our experimental setup is functionally the same as that described previously. \cite{dav.08a} To probe the Faraday rotation, we cool liquid $^3$He to $\sim600$ $\mu$K in an acoustic cavity formed by a transducer and a quartz reflecting plate and then slowly decrease the pressure in the cell from $\sim6$ to 3 bar. This pressure change, as well as an associated temperature change, continuously alters the frequencies of pair-breaking and the ISQ relative to a fixed transducer frequency of 88 MHz. Accordingly, the sound frequency passes through the ISQ and approaches pair-breaking along a trajectory similar to the red curve in Fig. \ref{fig:EnDiag}. As the difference between $\omega$ and $\Omega$ increases with decreasing pressure, both the transverse sound velocity, $c_t=\omega/q$, and the wavelength decrease, changing the standing wave condition in the acoustic cavity. This produces the high-frequency oscillations shown in the inset of Fig. \ref{fig:EnDiag}. As previously discussed, application of a magnetic field along the direction of sound propagation rotates the sound polarization, and this is seen as a modulation of the acoustic cavity oscillations, shown by the low-frequency envelope in the inset of Fig. \ref{fig:EnDiag}. Both effects are described by
\begin{equation}
V_Z\propto \cos\theta\sin\Big(\frac{2d\omega}{c_t}\Big), \label{transig}
\end{equation}
where $V_Z$ is the detected transducer voltage, $\theta$ is the angle of the sound polarization relative to the direction in which sound was generated, and $d=31.6\pm0.1$ $\mu$m is the cavity spacing.\cite{dav.08a}

In order to convert an acoustic trace into a form that can be related to the dispersion relation, Eq. \ref{fulldisp}, we first apply Eq. \ref{transig} to extract $\theta$ and $c_t$. From the sinusoidal dependence of $V_Z$ on $\theta$, we identify minima in the envelope as the polarization rotation angles $\theta=n\times90^\circ;\:n=1,3,5\dots$, and calculate intermediate angles from the modulation. Also from $V_Z$, we measure the period of the high-frequency oscillations,
\begin{equation}
1 \text{ Period}=2d\frac{\omega}{2\pi}\Big|\frac{1}{c_{t}}-\frac{1}{c_{ti}}\Big|, \label{dcper}
\end{equation}
that results from the change of the sound velocity.  In order to determine $c_t$ we use Eq. \ref{fulldisp} to calculate an initial value of the velocity, $c_{ti}$, near resonance, where the theoretical dispersion is accurate, and then use Eq. \ref{dcper} at higher frequencies, $\omega>\Omega$.

\section{Results and Discussion}
\subsection{Data}
Our $\theta$ and $c_t$ data are displayed in Fig. \ref{fig:data}. The abscissa in Fig. \ref{fig:data}(a), (c), and several of the following figures is the normalized difference in the square of the frequencies, $(\omega^2-\Omega_0^2)/\omega^2$.  In the following we will refer to this as the relative frequency shift, or just the shift. We use this scaling in preference to more direct variables such as pressure or temperature because it is a more explicit measure of changes in the dispersion that take place as the pressure or temperature change.  During a typical pressure sweep the temperature also increases slightly and both of these dependencies are reflected in $\Omega(P,T)$.\cite{dav.06} The (b) and (d) panels in Fig. \ref{fig:data} are the values of $\theta$ and $c_t$ at specific shifts versus magnetic field, taken as vertical cuts indicated by the dashed lines in panels (a) and (c). It is immediately clear that the transverse sound velocity is relatively insensitive to the magnetic field while the Faraday rotation angle, $\theta$ in Fig. \ref{fig:data}(b), is predominantly linear in field at low fields, but becomes nonlinear at higher fields. Additionally, there is a substantial decrease in the linear field term with increasing $(\omega^2-\Omega_0^2)/\omega^2$ which we find to be inconsistent with the theory of the dispersion \cite{sau.00a} expressed by Eqs. \ref{fulldisp} and \ref{linisq}, which we note was formulated only for the region of vanishing shift.

For our analysis, we exclude data below $(\omega^2-\Omega_0^2)/\omega^2=0.035$, which is represented by the left-most grey vertical line in Fig. \ref{fig:data}(a). This corresponds to the shift below which the highest field data is unreliable. As the sound frequency approaches the mode, the rotation angle diverges, and past a certain point the higher field data cannot be accurately determined. Restricting our analysis to shifts above this region ensures that our results are unaffected by this issue.

\subsection{Dispersion}
The dispersion, in the form attained by combining Eqs. \ref{fulldisp} and \ref{linisq}, can be solved to produce the Faraday rotation angle, given the temperature, pressure, and $g$. If $g$ were independent of shift the $\theta$ values calculated from the dispersion would show a decrease in the linear field dependence with increasing shift, as shown by the relative slopes of the solid lines in Fig. \ref{fig:data}(b), but the magnitude of that decrease is less than two thirds that of our data. However, we must also allow for the fact that the theory of Sauls and Serene \cite{sau.82} predicts that $g$ depends weakly on both $T$ and $P$.  For the range of our data, $P\sim6$\,$\rightarrow$\,$3$ bar and $T/T_c\sim0.47$\,$\rightarrow$\,$0.64$, the maximal expected change is $\delta g\sim+0.008$, which widens the discrepancy between data and theory even further by $\sim 13\%$. Therefore, the $(\omega^2-\Omega_0^2)/\omega^2$ dependence of the Faraday rotation angle we measure is incompatible with the theory of the transverse sound dispersion, Eqs. \ref{fulldisp} and \ref{linisq}, which was formulated for the near vicinity of the collective mode.\cite{sau.00a}

The experiments by Movshovich \et were performed at crossing, $\omega=\Omega$, and so they measured the field dependence directly.\cite{mov.88} In contrast, the transverse sound experiments in our work, as well as those of Davis {\it et al.},\cite{dav.08a} explore the full region of frequency between $\Omega$ and $2\Delta$, where the dispersion relation, Eqs. \ref{fulldisp} and \ref{linisq}, appears to be inapplicable.

To provide a framework for analysis we take a phenomenological approach making an assumption that the denominator on the right-hand side of the dispersion can be expanded in orders of field including terms up to $H^3$.  We modify Eq. \ref{linisq} to be of the form
\begin{IEEEeqnarray}{rCl}
&&\omega^2-\Omega_0^2-\frac{2}{5}q^2v_F^2-m_JA\gamma_{\text{eff}}H \nonumber\\ &&-m_J^2B\gamma_{\text{eff}}^2H^2-m_J^3C\gamma_{\text{eff}}^3H^3, \label{ABC}
\end{IEEEeqnarray}
where the terms containing $A$, $B$, and $C$ describe linear, quadratic, and cubic magnetic field dependences, respectively, and depend on frequency shift determined directly from experiment. As transverse sound couples only to the $m_J=\pm1$ substates, the linear and cubic terms switch sign for different substates, while the quadratic term is always negative since $m_J^2=1$. Within this framework, our choices for $m_J$ are consistent with the theoretical field dependence for $\Omega(H)$.\cite{sch.83,fis.86}

\subsection{Analysis}
We can relate both Faraday rotation angle and sound velocity data to the modified dispersion of the ISQ found by inserting Eq. \ref{ABC} into Eq. \ref{fulldisp}. It is helpful to use the following relations: \cite{sau.00a,sau.00b}
\begin{IEEEeqnarray}{rCl}
\theta&=&2d\,\delta q, \label{theta}\\
c_t&=&2\omega/(q_++q_-), \label{ct}
\end{IEEEeqnarray}
where $\delta q=|q_+-q_-|/2$, and $q_\pm$ is obtained by solving Eq. \ref{fulldisp} for $q$, setting $m_J=\pm1$. Because $c_t$ is inversely proportional to the average of $q_\pm$, its dependence on linear and cubic field terms cancels. Thus, the sound velocity depends predominantly on the quadratic field term. Conversely, $\theta$ depends most strongly on the linear and cubic terms since the quadratic term is suppressed in the difference between $q_+$ and $q_-$.
\begin{figure}[t!]
\includegraphics[width=3.5in]{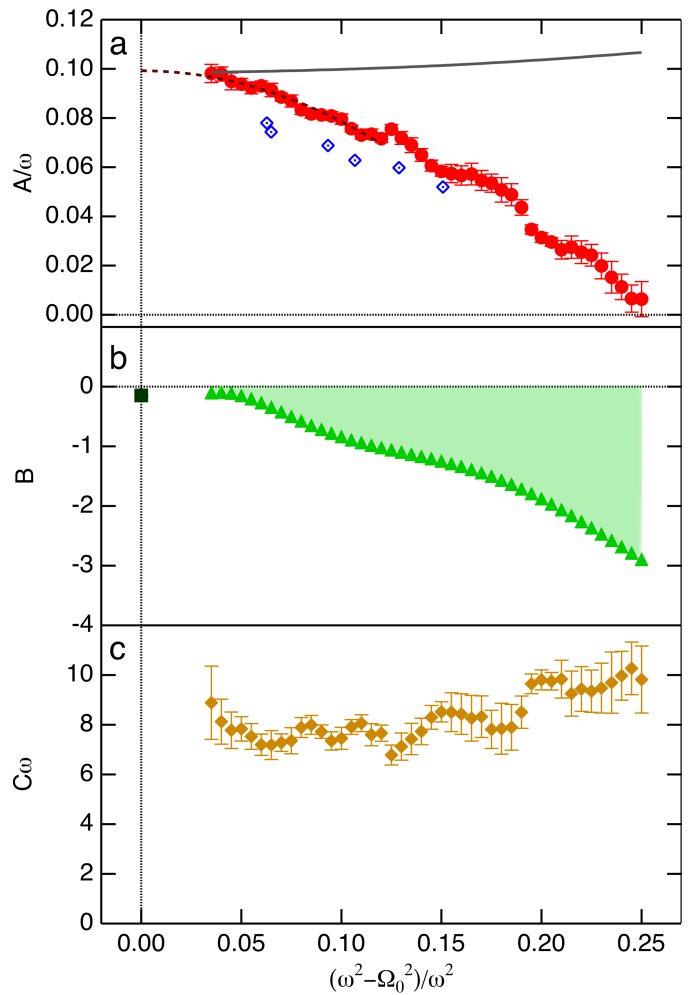}
\caption{\label{fig:ABC}Field dependence parameters $A$, $B$, and $C$ for Eq. \ref{ABC}, normalized to appropriate orders of $\omega$ to make them dimensionless, plotted against shift. (a) $A/\omega$ quantifies the linear field dependence, with the results from this work given by red circles, and data from Davis \et \cite{dav.08a} at the fixed pressure of $P=4.7$ bar shown as open blue diamonds, which we have reanalyzed as discussed later in the text. At low shift the dominant contribution to $A$ is from the Zeeman splitting of the ISQ for which the theory\cite{sau.00a} should be applicable. This splitting is predicted\cite{sau.82} to change slightly with $T$ and $P$ for a typical pressure sweep as shown by the solid grey curve. Clearly there are important contributions to $A$ for frequencies well away from the ISQ mode not described by the theory. The dark red dashed line indicates the extrapolation of our data to $\omega=\Omega_0$, where the theory is valid. (b) $B$ identifies an upper bound on the quadratic field dependences, green triangles, with the shaded area showing the possible range of the magnitude of B. The result from Movshovich \et \cite{mov.88} is shown as a black square at $\omega=\Omega_0$, consistent with our analysis. (c) The cubic field parameter, $C\omega$, is shown as solid orange diamonds.}
\end{figure}

Examination of the data in Fig. \ref{fig:data}(d) shows that $c_t$ varies little, if at all, with field. To quantify this we separate all the $c_t$ data in Fig. \ref{fig:data}(c) into bins of width $(\omega^2-\Omega_0^2)/\omega^2=0.005$, and fit the data in each bin to Eq. \ref{ct} as a function of field, using Eq. \ref{ABC} with $B$ as the only free parameter and setting $A=C=0$.\cite{col.12a} Due to the apparent field independence we can at best establish an upper bound for the quadratic dependence of $c_t$ on field, shown by green triangles in Fig. \ref{fig:ABC}(b), where the green shaded area represents the possible magnitude of $B$.

\begin{figure}[t!]
\includegraphics[width=3.5in]{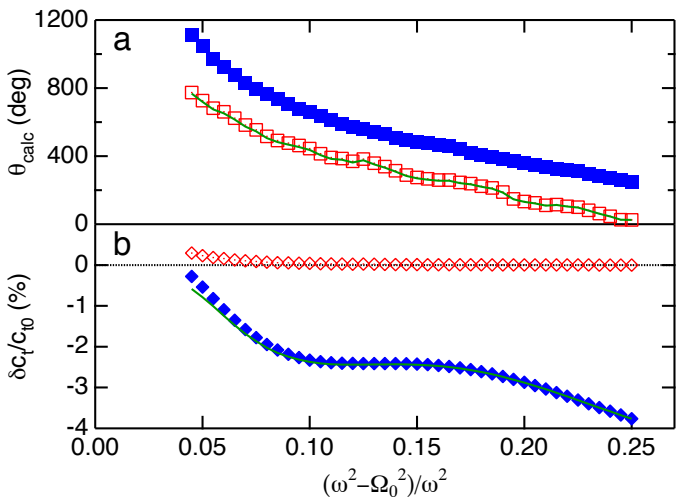}
\caption{\label{fig:thct}A comparison of the relative importance of linear, quadratic, and cubic field dependences on, a) $\theta_{\text{calc}}$ and b) $\delta c_t/c_{t0}=(c_{t\text{calc}}-c_{t0})/c_{t0}$ at $H=0.1$ T as a function of shift. Here $\theta_{\text{calc}}$ and $c_{t\text{calc}}$ are the rotation angle and sound velocity calculated by solving the dispersion for $q_\pm$ and inserting the result into Eqs. \ref{theta} and \ref{ct}; $c_{t0}$ is calculated at zero field. These calculations use Eq. \ref{ABC} with only the linear term, red open symbols; the linear plus quadratic terms, green lines; and linear, quadratic, plus cubic field terms, blue solid symbols. Adding the quadratic dependence is seen to change $\theta_{\text{calc}}$ very little, and $\delta c_t/c_{t0}$ a significant amount. Note that the quadratic field dependence, the $B$ term, is just an upper bound from our measurement of the field dependence of $c_t$ dictated by the precision of our measurement. Adding the cubic dependence changes $\theta_{\text{calc}}$ significantly, and $\delta c_t/c_{t0}$ very little, apart from very near the ISQ.}
\end{figure}

Nonlinear magnetic field effects play a significant role in $\theta$, as seen in Fig. \ref{fig:data}(b). If we use the values for $B$ established as a bound on the quadratic terms, we find that the effect of a quadratic field dependence on $\theta$ is negligible. In order to describe the observed nonlinearity we must include the cubic field term in Eq. \ref{ABC} containing the coefficient $C$ in our $\theta$ fits. With this inclusion our model describes the data well. The best fit values for $A$, $B$, and $C$ are shown in Fig. \ref{fig:ABC}, normalized to appropriate orders of $\omega$ to render them dimensionless. The fitting is performed self-consistently with care to ensure that the initial sound velocity, constrained by the theory, is correctly represented, and to ensure that our $A$ values are unaffected by any uncertainty in the high field data near the mode. The solid grey curve in Fig. \ref{fig:ABC}(a) is an extrapolation of the theory\cite{sau.82,sau.00a} to frequencies well above the mode frequency, outside of its range of validity, which illustrates the discrepancy between our data and the theory. The relative importance of linear, quadratic and cubic terms in Eq. \ref{ABC}, {\it i.e.} $A$, $B$, and $C$, on calculated values of $\theta$ and $c_t$ is displayed in Fig. \ref{fig:thct} and described in the caption.

\section{Nonlinear Field Dependence}
We can compare our results for a bound on $B$ with that of Movshovich \textit{et al.}\cite{mov.88} Their value for $B$ is presented as the black square in Fig. \ref{fig:ABC}(b). This was taken from the nonlinear effects seen in the $m_J=0$ substate, and analyzed assuming a field dependence of the form
\begin{equation}
\Omega(H)=\Omega_0+\alpha m_JH+\beta m_J^2H^2-\Gamma H^2, \label{movisq}
\end{equation}
leaving only the $\Gamma$ term to affect the field dependence of the $m_J=0$ state. Assuming that same form, our result combines the quadratic field terms, $B=\beta-\Gamma$, and so we can only say that their result appears to lie within the bound we have set from our measurement of the field dependence of the velocity of transverse sound. Their analysis did not include the possibility of a cubic field dependence, while ours yields a fairly constant value of $C$ across the entire relative frequency shift range.

\section{$g$-factor and the $f$-wave pairing strength}
\begin{figure}[t!]
\includegraphics[width=3.5in]{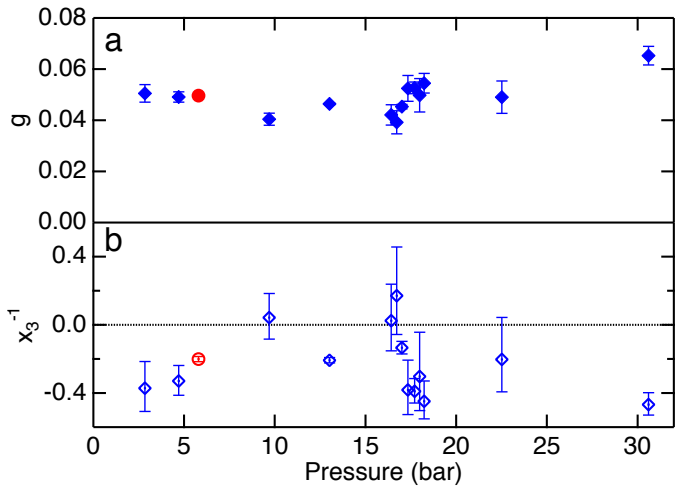}
\caption{\label{fig:x3s}Calculated $g$ and $x_3^{-1}$ results as a function of pressure. (a) Extrapolated zero-shift $g$ values for this work (solid red circle, error bars inside the data point) and reanalyzed data of Davis \et \cite{dav.08a} (solid blue diamonds). In the absence of Fermi liquid interaction effects, the expected weak-coupling value of $g$ for these temperatures and pressures varies between 0.033 and 0.04, well below most of these results.\cite{sau.82} (b) $f$-wave pairing parameter values calculated from $g$ values in (a). Data from this work (open red circle) and Davis \et (open blue diamonds) are presented. Positive values correspond to repulsive interactions, and negative values to attractive interactions; the data generally indicates attractive $f$-wave pairing.}
\end{figure}

We can use our results for $A$ in the long-wavelength limit to determine the $g$-factor of the ISQ. As the theoretical predictions \cite{sau.00a,sau.82} for both the dispersion and $g$ were made for $\omega\sim\Omega$, our data cannot be used directly to calculate $g$. However, as our sound frequency approaches $\Omega$, our $A$ values smoothly approach a limiting value $A_0$, as they are expected to in the region where the theory is robust. We use this observation to extrapolate our data to $\omega=\Omega_0$ in order to compare with the theory, which we accomplish by fitting the points closest to the mode to a quadratic, shown by the dark red dashed line in Fig. \ref{fig:ABC}(a), and taking the intercept, $A_0/\omega=0.0992$. Doing so we obtain \gtwo, shown as a solid red circle in Fig. \ref{fig:x3s}(a).

Previous measurements of $g$ for the ISQ have been reported. Using the acoustic Faraday effect, Lee \et \cite{lee.99} found $g=0.02\pm0.002$ at $P=4.32$ bar, and Davis \et \cite{dav.08a} measured $g$ at pressures from $\sim3-31$ bar. An error was made in the original calculations of Davis \textit{et al.}, later corrected,\cite{dav.11} but there was also a fundamental problem underlying the analysis which we have revised in the present work. We refer to this as reanalyzed data in Figs. \ref{fig:ABC} and \ref{fig:x3s}(a). Davis \et extrapolated their data to $T=0$ in order to avoid a region where $g$ exhibited an unexpected temperature dependence, which disagreed with the predictions of Sauls and Serene.\cite{sau.82} Upon our further investigation, we have found that this temperature dependence is actually the same $(\omega^2-\Omega_0^2)/\omega^2$ dependence in the linear magnetic field term, as shown in Fig. \ref{fig:data}(b), that falls outside the range of validity of the theory. This can be seen in the Davis \et data for 4.7 bar, shown as blue diamonds in Fig. \ref{fig:ABC}(a), which we have reanalyzed using the phenomenological dispersion described above. After reanalyzing their data for all pressures, we extrapolate to $\omega=\Omega_0$ to get the $g$ values shown by solid blue diamonds in Fig. \ref{fig:x3s}(a). These extrapolations are done based on a linear rather than quadratic fit, due to the limited amount of data available at each pressure as seen by the small number of blue diamonds in Fig. \ref{fig:ABC}(a).

The precise determination of the $g$-factor of the ISQ has impact beyond understanding the Zeeman splitting of the mode, since the $g$-factor is sensitive to $f$-wave pairing interactions. We use the parameter $x_3^{-1}\equiv1/\ln(T_3/T_c)$ to quantify the strength of these interactions, where $T_3$ would be the transition temperature for pairing in the $l=3$ angular momentum channel in the absence of other interactions. Negative values of $x_3^{-1}$ correspond to an attractive interaction.\cite{sau.82,hal.90} Using the theory of Sauls and Serene\cite{sau.82} we find \xthree, shown as an open red circle in Fig. \ref{fig:x3s}(b), giving $T_3/T_c\simeq0.007$ at $P=6$ bar. This result was calculated using the Fermi liquid parameter $F_2^s$ cited by Halperin and Varoquaux.\cite{hal.90} There are uncertainties in all the Fermi liquid parameters required for the analysis, and $F_2^s$ is not well known at low pressure; changing $F_2^s$ by $\pm0.2$ causes a change in $x_3^{-1}$ of $\mp0.107$.

Previous experiments have been interpreted in terms of $x_3^{-1}$ at various pressures. The zero-field frequencies of both the ISQ and another collective mode with $J=2$, the real squashing mode (RSQ), were predicted to depend on $f$-wave interactions,\cite{sau.81a} as was the magnetic susceptibility.\cite{fis.86,fis.88b} For pressures around 6 bar, $x_3^{-1}$ calculated from the RSQ frequency is $-0.06$,\cite{fra.89,hal.90} and two different ISQ frequency measurements gave $x_3^{-1}$ to be $-0.14$,\cite{mei.87,hal.90} and $0.025$.\cite{dav.06} In addition to uncertainty in these values from the insensitivity of the zero-field mode frequency to $x_3^{-1}$, they also contain uncertainties from the Fermi liquid parameters $F_2^a$ and $F_2^s$, such that a change in $F_2^a$ or $F_2^s$ of 0.2, within the uncertainty of the parameters, causes a change in $x_3^{-1}$ of about 0.05. Susceptibility measurements,\cite{hoy.81} at less than 1 bar, have been interpreted\cite{fis.88b} to give $x_3^{-1}=-1.75\pm0.15$.

In direct comparison with our data, previous $g$ measurements have been used to calculate $x_3^{-1}$. Sauls used the $g$ measurement of Lee \et \cite{lee.99} to calculate $x_3^{-1}\simeq-0.33$.\cite{sau.00a} We calculate $x_3^{-1}$ from the reanalyzed data of Davis \textit{et al.},\cite{dav.08a} shown in Fig. \ref{fig:x3s}(b) as open blue diamonds. While there is significant scatter, the general trend appears to agree with that of our data point. This result supports the identification of a recently discovered collective mode near pair-breaking as a $J=4^-$ mode, which relies on attractive $f$-wave interactions.\cite{dav.08b}

\section{Conclusion}
We have found significant nonlinear field effects in the dispersion relation for transverse sound in superfluid $^3$He. Theoretical predictions based on $qv_F/\omega<<1$ for the dispersion of transverse sound are applicable in a small frequency range above the mode frequency. Theoretical results over a wide frequency range with $qv_F/\omega\sim1$ are needed. We have introduced a model through which we have analyzed our data and quantified the field dependence of the dispersion up to cubic order. From the linear behavior, we determined the $g$-factor for the Zeeman splitting of the imaginary squashing mode, which implies a small but attractive $f$-wave pairing interaction at low pressure. Our result for the $f$-wave pairing interaction parameter, {\xthree} at $P=6$ bar, is in agreement with our reanalysis of the measurements of Davis \et \cite{dav.08a} for which the $l=3$ pairing channel is attractive at all pressures.

\section{Acknowledgments}
We would like to thank J.A. Sauls for his help with this work and A.M. Zimmerman for useful discussions, and acknowledge the support of the National Science Foundation, DMR-1103625.

\bibliography{refs}{}
\end{document}